\documentclass[aip,jcp,twocolumn,amsfonts,amsmath,amssymb,floatfix,10pt,reprint]{revtex4-1}
\usepackage{graphicx}
\usepackage{algorithm}
\usepackage{algpseudocode}
\usepackage{hyperref}

\usepackage{color}
\usepackage[normalem]{ulem}

\algnewcommand{\LineComment}[1]{\State\(\triangleright\) #1}

\DeclareMathOperator{\sqrtv}{sqrt}

\begin{document}
\title{Tensor-Structured Coupled Cluster Theory}

\author{Roman Schutski}
\affiliation{Department of Chemistry, Rice
University, Houston, Texas, 77251-1892, USA}

\author{Jinmo Zhao}
\affiliation{Department of Chemistry, Rice
University, Houston, Texas, 77251-1892, USA}

\author{Thomas M. Henderson}
\affiliation{Department of Chemistry,
Rice University, Houston, Texas, 77251-1892, USA}
\affiliation{Department of Physics and Astronomy, Rice University,
Houston, Texas, 77251-1892, USA}

\author{Gustavo E. Scuseria}
\affiliation{Department of Chemistry,
Rice University, Houston, Texas, 77251-1892, USA}
\affiliation{Department of Physics and Astronomy, Rice University,
Houston, Texas, 77251-1892, USA} \date{\today}

\begin{abstract}
We derive and implement a new way of solving coupled cluster
equations with lower computational scaling. Our method is based 
on decomposition of both amplitudes and two electron integrals, 
using a combination of tensor hypercontraction and canonical polyadic
decomposition. While the original theory scales as $O(N^6)$ 
with respect to the number of basis functions, we demonstrate 
numerically that we achieve sub-millihartree difference from 
the original theory with $O(N^4)$ scaling. This is 
accomplished by solving directly for the factors that decompose
the cluster operator. The proposed scheme is quite general
and can be easily extended to other many-body methods.
\end{abstract}

\maketitle

\section{{Introduction}
\label{sec:introduction}}

Many basic building blocks of quantum theories are tensors.  Examples
include the one- and two-electron integrals defining the Hamiltonian
or the cluster operators of coupled cluster (CC) theory, which instead
define the wave function.  Unfortunately, algebraic manipulations with
tensors have a significant numerical cost, which tends to grow
exponentially with the dimension $d$ of the tensors and often makes
these manipulations the computational bottleneck of the theories.

The cost of tensor manipulations can be significantly reduced by some
form of tensor decomposition in which a $d$-dimensional tensor is
expressed in terms of lower dimensional objects.  For example, the
resolution of identity (RI, see Ref. \onlinecite{weigend2009approximated}
and references therein)  can be used
to decompose the two-electron integrals.  More recently, the tensor
hypercontraction\cite{hohenstein_thc1,hohenstein_thc2,hohenstein_thc3,
kokkila2015tensor,parrish2014communication, hohenstein2013tensor}
(THC) scheme of Hohenstein, Parrish, and Martinez has been introduced.
There, a fourth-order tensor is represented by a contraction of five
factor matrices, some of which can be the same if one wants to enforce
symmetries of the original tensor.  Related to THC is the canonical
polyadic decomposition\cite{hitchcock1927expression,de2006link} (CPD),
which as we will explain later can be regarded as its building block.

These tensor decompositions have been used in various ways to
introduce low-scaling versions of conventional electronic structure
methods. Tensor hypercontraction has been applied by Hohenstein and
Kokkila\cite{hohenstein_cc2} to the CC2 method, where it was used to
represent an electron interaction potential. Shenvi \emph{et al.} did
the same in their Reduced Density Matrix algorithm.\cite{shenvi}
Benedict \emph{et al.} used polyadic decomposition of amplitudes and
electron interaction integrals in the coupled cluster
doubles and full configuration interaction methods.\cite{benedict_mp2,
benedict_ccd} While working on this manuscript we also learned about
the recent work of Hummel \emph{et
al.},\cite{gruneis} who showed that by using THC of the electron
interaction in the context of the distinguishable cluster doubles or
linearized coupled cluster singles and doubles methods, one can
achieve a reduction of the computational cost from $O(N^6)$ to
$O(N^5)$, where $N$ is the number of basis functions.

Here we apply tensor decompositions based on the THC to coupled
cluster with single and double excitations
(CCSD).\cite{purvis1982full, scuseria_ccsd} The cost of the original
CCSD scales as $O(N^6)$,
but by using tensor decomposition we can reduce the cost to scale as
$O(N^4)$. In most previous applications, THC was used to decompose the
electron repulsion integrals, and grids in real space were employed to
build the decomposition.  We show how to build the THC algebraically
for the full fourth-order tensor in $O(N^5)$ cost, or $O(N^4)$ cost if
the resolution of identity\cite{ahmadi1995coulomb} is employed, and
compare different ways of doing so. We also explain how to optimize
all factors of the THC in $O(N^4)$ cost when solving iterative
equations with decomposed tensors, such as in the CCSD
method. By optimizing all factors of the THC, our
implementation achieves the same $\sim 0.5$ millihartree accuracy as
previous work\cite{hohenstein_thc3} which used the THC but with ranks
which are roughly half as large. However, we should emphasize that our
method is general and is not limited to THC; rather, it can be used
with \textit{any} suitable tensor decomposition.

\section{{Notation and Terminology}
\label{sec:notation}}
Throughout this work we will use
notation and diagrams which are common in the literature of tensor
decompositions but which may be unfamiliar to the quantum chemistry
community.  A short review of our diagrammatic notation is available
in Appendix \ref{sec:appendix_diagrams}; we summarize our notation and
terminology here.

One of the most basic properties of a tensor $T$ is its
\textit{order}, which is just its dimensionality and corresponds to
the number of indices in its basis representation.  Thus, a four-index
object (if a tensor) corresponds with a fourth-order tensor.  We
sometimes refer to a first-order tensor as a vector and a second-order
tensor as a matrix.  Generically we denote matrices and tensors by
capital letters, and vectors by boldface lower-case letters.

The \textit{rank} of a tensor is the dimension of the auxiliary
indices used in a particular tensor decomposition. As there are a
great variety of tensor decompositions, the rank of a high-order
tensor is not defined as strictly as in the case of matrices, and may
consist of one number or a set of numbers; for our purposes, if the
tensor has more than one rank it is convenient to require all its
ranks to be equal.  Different definitions of tensor rank have
significantly different properties; for more information consult the
review of Kolda \emph{et al.}\cite{kolda2009tensor}.  It should be
clear from context what dimensions are meant in each particular case
in the text.

The Frobenius norm of a tensor $T$ is denoted as $\|T\|$ and is given
by
\begin{equation} \| T \| = \sqrt{\sum_{pqrs\ldots}
T_{pqrs\ldots} \, T_{pqrs\ldots}^\ast}
\end{equation} where the superscript $\ast$ denotes complex
conjugation; thus, it is simply the square root of the sum of the
squares of the tensor's entries.

We will require a few kinds of tensor product in this work.  The
Kronecker product is written as $\otimes$, and is defined via
\begin{equation} C = A \otimes B \Leftrightarrow C_{rp, sq} = A_{p,q}
\cdot B_{r,s}.
\label{eq:kron_product}
\end{equation} It is also convenient to introduce a column-wise
Kronecker product known as the Khatri-Rao\cite{liu2008hadamard}
product; this is denoted by $\odot$ and is defined as
\begin{equation} D = A \odot B ~~\Leftrightarrow ~~D_{qp,\alpha} =
A_{p,\alpha} \cdot B_{q,\alpha}.
\label{eq:khatri_rao_product}
\end{equation}

In the foregoing and throughout this manuscript, indices $p, q, r, s,
\ldots$ correspond to general orbital labels and Greek letters
$\alpha, \beta, \gamma, \ldots$ denote indices of the CPD, THC, and
singular value decompositions.  We follow the traditional notation
that the indices $i,j,k,l\ldots$ represent occupied orbitals
specifically, while $a,b,c,d\ldots$ represent virtual orbitals.  We
also use composite indices such as $pq$ which are defined as
\begin{equation} pq \equiv p + \dim(\{p\})\cdot (q - 1).
\label{eq:combined_indices}
\end{equation}
Curly braces $\{\}$ denote sets and $\dim()$ means the
number of elements in the set.

Finally, summation is impled for any indices which appear more than
once within a product.  The transpose of a matrix $M$ is $M^T$ and its
inverse is $M^{-1}$; if $M$ is singular or not square $M^{-1}$ refers
to the pseudoinverse\cite{barata2012moore} of $M$.  We will use
$\sqrtv()$ for the element-wise square root operation:
\begin{equation}
  \sqrtv(M)_{pq} = \sqrt{M_{pq}}.
\end{equation}
\section{{Tensor Decompositions}
\label{sec:tensor_decompositions}} The tensor hypercontraction
decomposition can be regarded as a combination of two well established
factorizations: a rank decomposition of a matrix such as the
eigenvalue or singular value decomposition (SVD) on the one hand and
the canonical polyadic decomposition\cite{hitchcock1927expression,
kolda2009tensor, de2006link} of third order tensors on the other.
Thus, we first discuss these two ideas.

\subsection{{Resolution of Identity and Singular Value Decomposition}
\label{sec:ri_singular_value}}
The computation of a rank-revealing
decomposition for the electron interaction tensor is well studied and
is known by the names of the resolution of identity (RI) or density
fitting.\cite{vahtras1993integral, boman2008method, sierka2003fast}
By introducing an
auxiliary basis, the Coulomb interaction can be written as a
contraction of three tensors
\begin{equation} V_{pqrs} \approx U_{pq}^{\alpha}
D_{\alpha,\alpha^{\prime}} \tilde{U}_{rs}^{\alpha^{\prime}},
\label{eq:ri_decomposition}
\end{equation}
where $V$ is the Mulliken-ordered interaction, $U$ and
$\tilde{U}$ are (possibly different) three index integrals, and $D$ is
a generalized overlap.\cite{ahmadi1995coulomb} Diagrammatically the
same expression is
\begin{equation}
\vcenter{\hbox{\includegraphics[height=12mm]{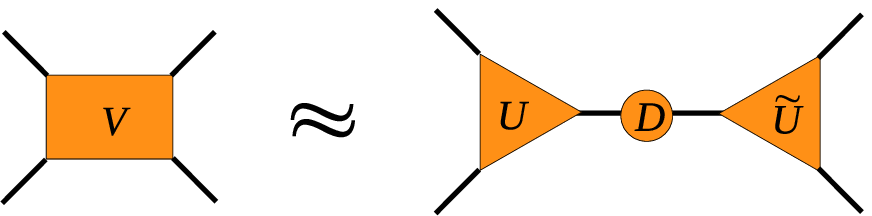}}}.
\label{fig:ri_decomposition}
\end{equation}
As one may see, RI has the same basic form of a
singular value or an eigenvalue decomposition of the interaction
tensor.  It is known that the error in the RI approximation of the
Coulomb operator decays exponentially with the auxiliary basis size
$r_\mathrm{RI} = \dim(\{\alpha\})$, and negligible errors can be
reached with the number of auxiliary basis functions scaling as
$O(N)$.\cite{sierka2003fast}

We note that for a given rank $r_\mathrm{RI}$, the lowest error RI
decomposition can be calculated using the singular value decomposition
of the matrix $V_{pq,rs}$ and taking $D$, $U$, $\tilde{U}$ to be the
singular values, left, and right singular vectors, respectively.  The
optimality of the factorization will then be guaranteed by the
Eckart-Young theorem.\cite{eckart1936approximation} Although this
approach is not generally used for practical calculations due to its
computational cost, which scales as $O(N^4 \cdot r_\mathrm{RI})$, we
employed it in some of our test calculations. We also note that a
popular practical option in the case of two electron integrals is the
use of Cholesky
decomposition,\cite{koch2003reduced,harbrecht2012low,weigend2009approximated}
but this method is limited to symmetric tensors only.

We have said that $V$ is the Mulliken-ordered interaction tensor.  The
restriction to Mulliken ordering is important, because the order of
indices in the original tensor $V_{pq, rs}$ crucially influences the
size of the rank $r_\mathrm{RI}$ for a fixed approximation error.
Indeed, while the SVD of the Mulliken-ordered electron interaction
$V_{pq, rs}$ yields $O(N)$ non-zero singular values, the matrix
$V_{pr,qs}$ formed from a Dirac-ordered interaction tensor has
$O(N^2)$ non-zero singular values.  This explains why there is no
practical RI-like approximation for Dirac-ordered two-electron
integrals (or, equivalently, exchange contribution in
the context of the Hartree-Fock method).

The RI decomposition itself can readily lead to reduced scaling of
some quantum chemistry algorithms. If the contractions of the electron
interaction involve mostly indices $p,q$ and $r,s$, but not cross
combinations between them (e.g. contractions where one tensor has
indices $pq$ and $rs$ while another has indices $pr$ and $qs$), then a
reduction of cost can be achieved, such as in the RI-MP2
approach.\cite{ayala1999linear, werner2003fast,
izmaylov2008resolution} When these cross combinations occur, however,
one needs to search for additional structure in the operator
tensors. The latter can be achieved by the canonical polyadic
decomposition.

\subsection{{Canonical Polyadic Decomposition}
\label{sec:canonical_polyadic_decomposition}}
A polyad is a rank one
tensor, expressible, for example, by
\begin{equation}
X_{ijk\ldots} = a_i \, b_j \, c_k \, \ldots
\end{equation}
or more abstractly as a series of Kronecker products
\begin{equation}
X = \ldots \otimes c \otimes b \otimes a.
\end{equation}
Note that we multiply factors in inverse order; this is
simply to preserve a consistent column-major indexing of tensors.

A polyadic decomposition of a tensor is thus a decomposition as a sum
of polyads:\cite{de2006link}
\begin{equation}
T_{pqr\ldots} = \sum_\alpha a_p^\alpha \, b_q^\alpha \,
c_r^\alpha \ldots
\end{equation}
or, more abstractly,
\begin{equation} T = \sum_\alpha \ldots \otimes \boldsymbol{c}_\alpha
\otimes \boldsymbol{b}_\alpha \otimes \boldsymbol{a}_\alpha.
\label{eq:cpd_definition}
\end{equation}
The canonical polyadic decomposition is the polyadic
decomposition of lowest rank.  It may be seen as one of the
generalizations of SVD to third and higher order tensors, and for
dimensions greater than 2, the CPD is unique under mild
conditions.\cite{kruskal1977three, sidiropoulos2000uniqueness}

As can be seen from the definition of Eq.~\ref{eq:cpd_definition},
some matrix factorizations (for example, QR or LU factorizations) can
be thought of as a CPD of matrices.  In dimensions greater than 2,
however, no closed form algorithm to extract the CPD is known, and one
must rely on iterative optimization techniques to approximate the
CPD.\cite{sorber2013optimization} Substantial effort has been made by
the mathematical community to develop approaches for doing so.
Typical algorithms are the alternating least
squares\cite{comon2009tensor} (ALS), gradient descent by means of, for
example, the method of Broyden, Fletcher, Goldfarb, and Shanno (BFGS),
and nonlinear least squares (NLS)
methods.\cite{sorber2013optimization} We refer the reader to the
corresponding reviews\cite{kolda2009tensor, sidiropoulos2016tensor}
for further details.  We have used the
Tensorlab\cite{vervliettensorlab} program by Lathauwer \emph{et al.}
for calculating the CPD in this work.

The polyadic decomposition can be expressed more conveniently through
the Khatri-Rao product.  If the vectors $\boldsymbol{a}$,
$\boldsymbol{b}$, and $\boldsymbol{c}$ of Eq.~\ref{eq:cpd_definition}
are stacked together as columns of matrices $A = \{ \boldsymbol{a}
\}$, $B = \{ \boldsymbol{b} \}$, $C = \{\boldsymbol{c}\}$, then the
polyadic decomposition can be written as
\begin{equation} T_{pqr} = ((B \odot A) \cdot C^{T})_{pqr},
\end{equation} which diagrammatically is
\begin{equation}
\vcenter{\hbox{\includegraphics[height=16mm]{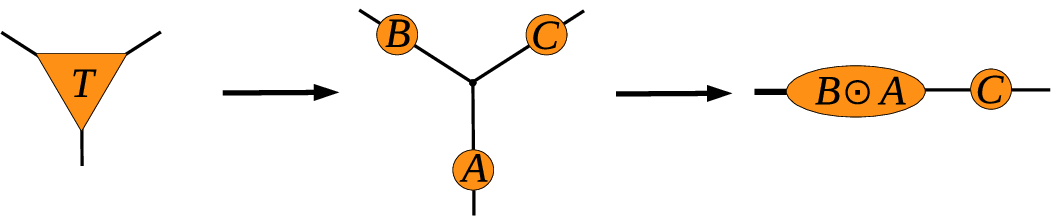}}}.
\label{fig:cp_decomposition}
\end{equation}

\subsection{{Tensor Hypercontraction}
\label{sec:tensor_hypercontraction}}
The THC is a factorization of
fourth-order tensors and can be seen as a combination of a singular
value decomposition and a canonical polyadic decomposition. The THC
approximation can be written as
\begin{equation}
\begin{split} V_{pqrs} & = W^{1}_{p,\alpha} W^{2}_{q, \alpha}
X_{\alpha, \beta} W^{3}_{r, \beta} W^{4}_{s, \beta} \\ &= ((W^{2}
\odot W^{1}) \cdot X \cdot (W^{4} \odot W^{3})^{T})_{pqrs}.
\end{split}
\end{equation}
The THC can be viewed as a further approximation over
RI, which is clear from the following diagram:
\begin{equation}
\vcenter{\hbox{\includegraphics[width=0.8\columnwidth]{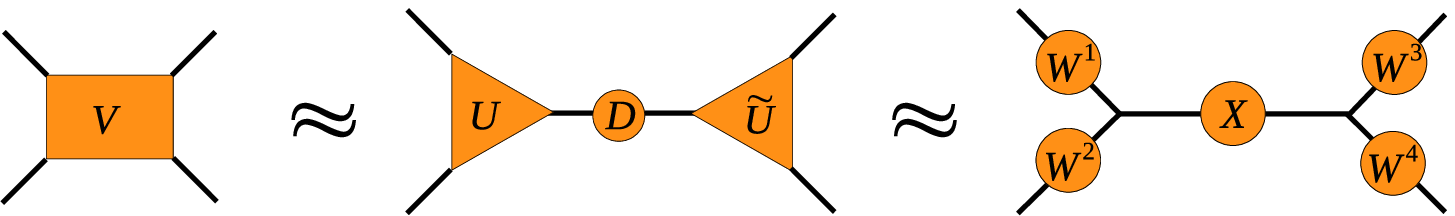}}}.
\label{fig:ri_cpd}
\end{equation}
The sizes of the auxiliary indices $\alpha$ and $\beta$
are the ranks of the decomposition.  In all subsequent expressions
$r_\mathrm{THC} = \dim(\{ \alpha \}) = \dim(\{ \beta \})$ for
simplicity, although there is no fundamental restriction that
$\dim(\{\alpha\}) = \dim(\{\beta\})$.  Using the analogy with density
fitting, several authors\cite{hohenstein_thc2, gruneis} have
speculated that the optimal rank of THC scales as $r_\mathrm{THC} =
O(N)$ in the case of the electron interaction tensor.  We confirm this
numerically in Sec.~\ref{sec:tensor_structured_cc}.

\subsection{{Algorithms for Tensor Hypercontraction}
\label{sec:optimization_of_thc}}
\subsubsection{{Composite Method}
\label{sec:compositemethod}}
The diagram in Eq.~\ref{fig:ri_cpd}
suggests one possible way to calculate the THC of an order-4 tensor as
a combination of the singular value and canonical polyadic
decompositions. The following diagram depicts the procedure we call
THC-CPD:
\begin{equation}
\vcenter{\hbox{\includegraphics[width=0.85\columnwidth]{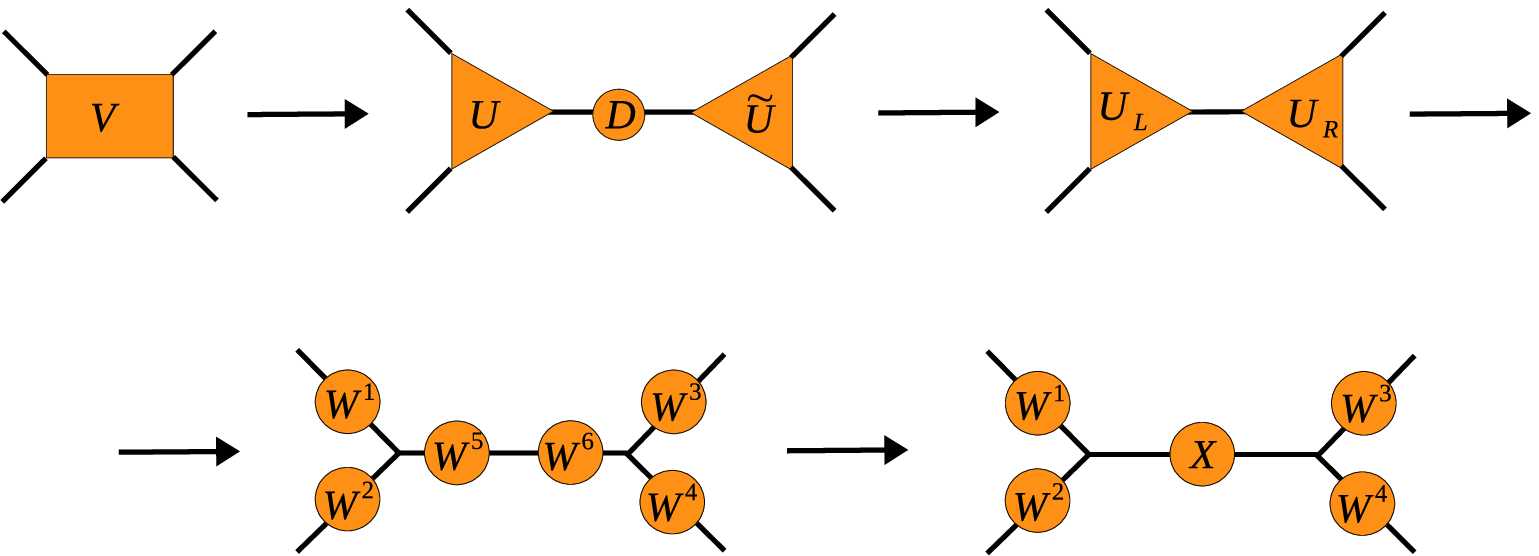}}}.
\label{fig:thc_cpd}
\end{equation}
First, one can reshape the original tensor $V$ with
dimensions $I_{1} \times I_{2} \times I_{3} \times I_{4}$ into a
matrix of shape $I_{1} I_{2} \times I_{3} I_{4}$ and apply a truncated
SVD of rank $r_\mathrm{SVD}$ to it. We chose to multiply square roots
of singular values into left and right singular vectors.  Note that
this produces matrices $U_{L}$ and $U_{R}$ of identical norm.  Next,
the left and right matrices of shapes $I_{1} I_{2} \times
r_\mathrm{SVD}$ and $I_{3} I_{4} \times r_\mathrm{SVD}$ are reshaped
into third-order tensors of shapes $I_{1} \times I_{2} \times
r_\mathrm{SVD}$ and $I_{3} \times I_{4} \times r_\mathrm{SVD}$,
respectively.  The CPD of rank $r_\mathrm{CPD}$ is calculated for each
of those tensors separately with any algorithm of choice, with each
algorithm limited to $n_{it}$ iterations.  Finally, those factors of
the CPD which do not have external indices can be merged into a single
factor $X$.

Algorithm~\ref{code:thc_cpd} summarizes the composite method we
employ, along with the computational scaling of its steps for a tensor
of size $N\times N \times N \times N$, where we used \emph{cpd()} to 
denote a CPD method of choice.
\begin{algorithm}[H]
  \caption{Computing the THC using CPD}\label{code:thc_cpd}
  \begin{algorithmic}[1]
  \Function{thc-cpd}{$V, r_\mathrm{SVD}, r_\mathrm{CPD}$}
  \State $I_{1}, I_{2}, I_{3}, I_{4} \gets$ size($V$)
  \State $V \gets$ reshape($V, I_{1}\cdot I_{2}, I_{3} \cdot I_{4}$)
  \State $U, D, \tilde{U} \gets$ svd($V, r_\mathrm{SVD}$)\Comment{$O(N^4
\, r_\mathrm{SVD})$}\label{code:thc_cpd:1}
   \State $U_L \gets U_L \cdot$ sqrt($D$)\Comment{$O(N^2 \, r_\mathrm{SVD}^2)$}
   \State $U_R \gets \tilde{U} \cdot $ sqrt($D^{\dagger}$)
   \State $W^1, W^2, W^5 \gets $cpd($U_L,
r_\mathrm{CPD}$)\Comment{$O(N^2 \, r_\mathrm{SVD} \, r_\mathrm{CPD} \,
n_{it})$}\label{code:thc_cpd:2}
   \State $W^3, W^4, W^6 \gets $cpd($U_R,
r_\mathrm{CPD}$)
    \State $X \gets W^5 \cdot$
${W^6}^{T}$\Comment{$O(r_\mathrm{CPD}^2 \cdot r_\mathrm{SVD})$}

  \Return $W^1, W^2, W^3, W^4, X$
  \EndFunction
  \end{algorithmic}
\end{algorithm}

A similar scheme was employed by Hohenstein \emph{et al.} in their
initial work on THC.\cite{hohenstein_thc1} The scaling of this
algorithm is dominated by the truncated SVD in
step~\ref{code:thc_cpd:1}. If the optimal rank of the SVD is of order
$O(N)$, the algorithm is of $O(N^5)$ cost if $r_\mathrm{SVD} = O(N)$
and $O(N^6)$ in the worst case.  The SVD can be avoided if substitute
singular vectors are available for the tensor $V$.  In the case of the
electron interaction, such substitutes are given by the 3-index
integrals coming from the RI approximation.  The auxiliary dimension
$r_\mathrm{RI}$ is of $O(N)$.

A faster Algorithm~\ref{alg:composite_ri} based on the RI approximation
can be formulated as
follows.  We start with third-order tensors $U$, $\tilde{U}$ of shapes
$I_1 \times I_{2} \times r_\mathrm{RI}$ and $I_{3} \times I_{4} \times
r_\mathrm{RI}$ respectively, and an overlap matrix $D$ of shape
$r_\mathrm{RI} \times r_\mathrm{RI}$ A matrix root $D^{\frac{1}{2}}$
of $D$ is calculated using the SVD or eigenvalue decomposition.  This
matrix is then multiplied into order-3 tensors $U$ and $\tilde{U}$,
which yields left and right third-order tensors $U_L$ and $U_R$.  If
the size of the RI basis is large and $U_L$ equals $U_R$, as in the
case of 3-index integrals, an optional compression step can be
applied (Algorithm~\ref{alg:ri_compression}): the auxiliary 
dimension of $U_L$ and $U_R$ is reduced by a
truncated SVD of rank $r_\mathrm{SVD}$.  Finally, a CPD of rank
$r_\mathrm{CPD}$ is calculated for the left and right parts, and the
resulting factors with no external indices are merged into a single
factor $X$. The resulting algorithm is listed below:
\begin{algorithm}[H]
  \caption{Computing the THC using CPD and RI}\label{code:thc_cpd_df}
  \begin{algorithmic}[1] 
  \Function{thc-cpd-ri}{$U, \tilde{U}, D,
r_\mathrm{SVD}, r_\mathrm{CPD}$}
  \State $Q \Lambda \tilde{Q} \gets $
svd($D$)\Comment{$O(r_\mathrm{RI}^3)$}
 \State $D^{\frac{1}{2}} \gets Q
\cdot $ sqrt($\Lambda$)
 \State $U_L \gets U \cdot
D^{\frac{1}{2}}$\Comment{$O(N^2 \, r_\mathrm{RI}^2)$}
 \State $U_R \gets \tilde{U} \cdot {D^{\frac{1}{2}}}^{\dagger}$
 \If{$U_R = U_L$}
 \State $U_R \gets $ compress($U_R, r_\mathrm{SVD}$)\Comment{Optional}
 \State $U_L \gets $ compress($U_L, r_\mathrm{SVD}$)\Comment{$O(N^2 \,
r_\mathrm{RI} \, r_\mathrm{SVD})$}
  \EndIf
   \State $W^1, W^2, W^5 \gets $cpd($U_L,
r_\mathrm{CPD}$)\Comment{$O(N^2 \, r_\mathrm{SVD} \, r_\mathrm{CPD} \,
n_{it})$}\label{code:thc_cpd_df:1}
   \State $W^3, W^4, W^6 \gets
$cpd($U_R, r_\mathrm{CPD}$)
   \State $X \gets W^5 \cdot$
${W^6}^{T}$\Comment{$O(r_\mathrm{CPD}^2 \cdot r_\mathrm{SVD})$}

  \Return $W^1, W^2, W^3, W^4, X$
  \EndFunction
  \end{algorithmic}
  \label{alg:composite_ri}
\end{algorithm}
\begin{algorithm}[H]
  \caption{Compressing the RI basis}\label{code:thc_cpd_df_compress}
  \begin{algorithmic}[1] \Function{compress}{$U, r_\mathrm{CPD}$}
\State $I_{1},I_{2},I_{3} \gets$ size($U$) \State $U \gets $
reshape($U, I_{1} \cdot I_{2}, I_{3}$) \State $Q A \tilde{Q} \gets $
svd($U, r_\mathrm{SVD}$)\Comment{$O(I_{1} \, I_{2} \, I_{3} \,
r_\mathrm{SVD})$} \State $U \gets Q A$\Comment{$O(I_{1} \, I_{2} \,
r_\mathrm{SVD}^2)$} \State $U \gets $ reshape($U, I_{1}, I_{2},
r_\mathrm{SVD}$) \State \textbf{return} $U$
    \EndFunction
  \end{algorithmic}
    \label{alg:ri_compression}
\end{algorithm}
The overall scaling of Algorithm~\ref{code:thc_cpd_df} may be
dominated either by the $O(N^2 \, r_\mathrm{RI}^2)$ cost of the SVD
and matrix multiplications or by the $O(N^2 \, r_\mathrm{SVD} \,
r_\mathrm{CPD} \, n_{it})$ cost of the iterative algorithm of the CPD.
In practical calculations we found that the latter contribution,
despite scaling mildly with the system size $N$ and optimal ranks
$r_\mathrm{CPD}$ and $r_\mathrm{SVD}$, is always dominant because of
the large number of iterations required by the CPD algorithm. This
motivated us to look for an equivalent algorithm to build the THC
decomposition directly.

\subsubsection{{Direct Method}
\label{sec:direct_method}}
We follow Sorber \emph{et
al.}\cite{sorber2013optimization} to build a simple alternating least
squares algorithm for THC.  We begin by introducing the approximation
of a fourth-order tensor $V$ by its THC decomposition $\tilde{V}$,
which we recall is
\begin{subequations}
\begin{align} \tilde{V}_{ijkl} &= W^{1}_{p,\alpha} W^{2}_{q, \alpha}
X_{\alpha, \beta} W^{3}_{r, \beta} W^{4}_{s, \beta} \\ &= [(W^2 \odot
W^1) \cdot X \cdot (W^4 \odot W^3)^T]_{ijkl}.
\end{align}
\end{subequations}
Then we can define an error tensor
\begin{equation}
\Delta_{V} = V - \tilde{V},
\end{equation}
whose Frobenius norm is just
\begin{equation} f = \|\Delta_V\|^2 = \left(V^\ast_{pqrs} -
\tilde{V}^\ast_{pqrs}\right) \, \left(V_{pqrs} -
\tilde{V}_{pqrs}\right).
\label{eq:cost_function}
\end{equation}
Diagrammatically, this is
\begin{equation}
\vcenter{\hbox{\includegraphics[height=38mm]{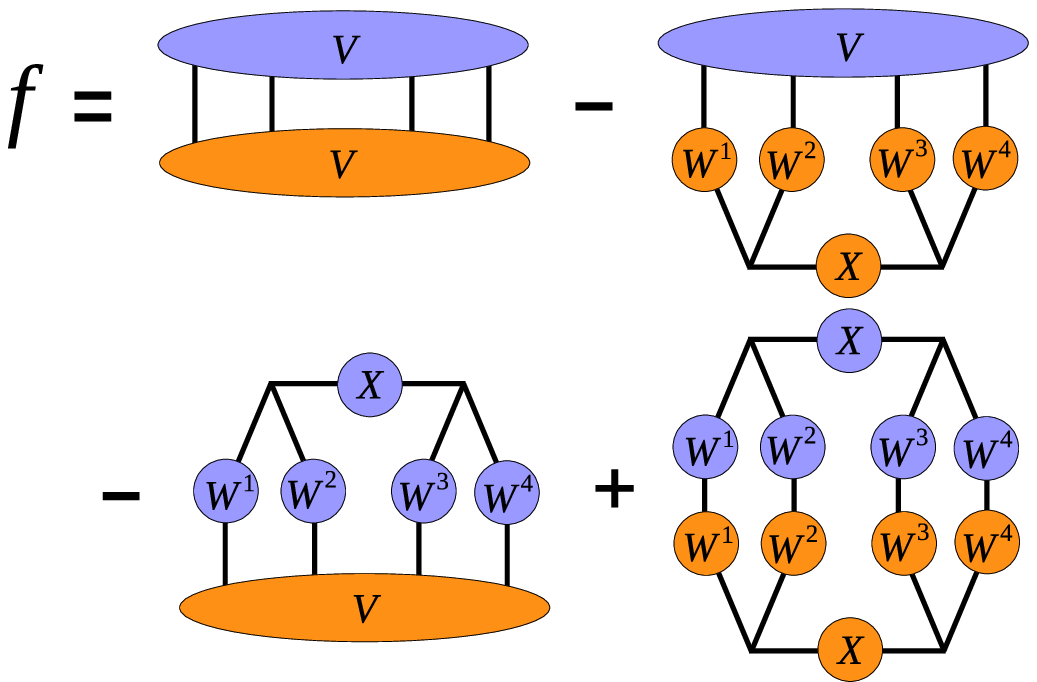}}}.
\label{fig:cost_function}
\end{equation}
Clearly, the best possible THC approximation to $V$
will correspond to a minimum of the cost function $f$.  We note that
$f$ is a real-valued analytic function, and hence $\frac{\partial
f}{\partial W} = (\frac{\partial f}{\partial W^{\ast}})^{\ast}$, where
$W \in \{W^1, W^2, W^3, W^4, X\}$.

In order to minimize the cost function, we proceed with the
calculation of its gradient, which can be easily done using
diagram~\ref{fig:cost_function}.  The partial derivative of $f$ with
respect to $W^1$ is
\begin{equation}
\vcenter{\hbox{\includegraphics[height=22mm]{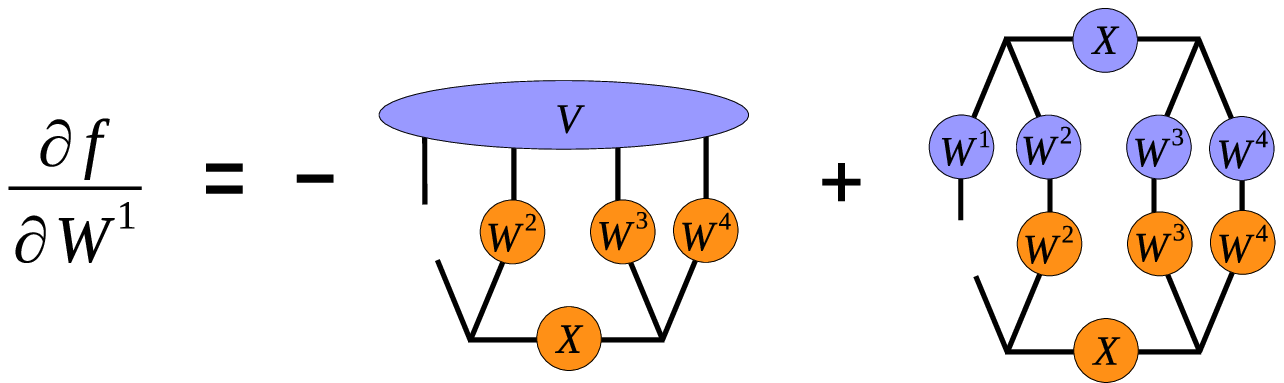}}}
\label{fig:cost_function_partial}
\end{equation}
and the full gradient of $f$ can be found in the
supplementary material.  Noting that $\frac{\partial f}{\partial W}$
is linear in $W^\ast$, we contract all factors around $W^\ast$ into an
environment matrix $A$, as seen in diagram~\ref{fig:least_squares_w1},
and set the derivative to zero:
\begin{equation}
\vcenter{\hbox{\includegraphics[height=24mm]{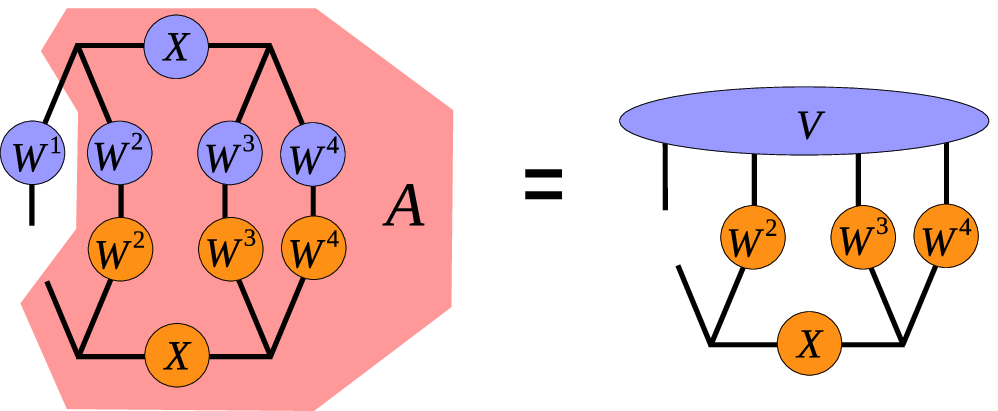}}}
\label{fig:least_squares_w1}
\end{equation}
We end up with a problem
\begin{equation}
A \cdot W^\ast = B.
\label{eq:least_squares_w1}
\end{equation}
The solution to Eq.~\ref{eq:least_squares_w1} can be
obtained by taking the inverse of $A$ (or a pseudoinverse, if $A$ is a
rank-deficient matrix).  The final expression for ${W^{1}}^{\ast}$ is
given diagrammatrically as
\begin{equation}
\vcenter{\hbox{\includegraphics[height=20mm]{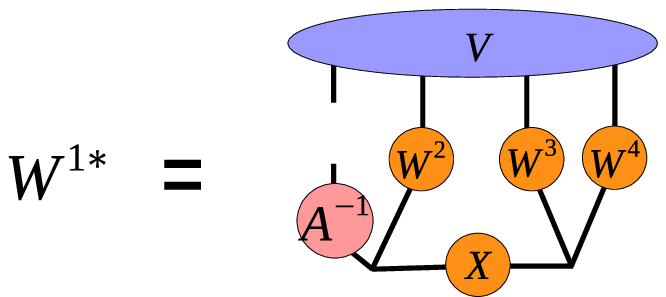}}}.
\label{fig:least_squares_w1_sol}
\end{equation}
As made clear by the diagram, if both ranks of the THC
decomposition are $r_\mathrm{THC}$, then the construction of the
environment matrix $A$ scales as $O(r_\mathrm{THC}^3)$, as does
computing its generalized inverse.  If each of the
dimensions of $V$ equals $N$, then the cost of calculating
${W^1}^{\ast}$ scales as $O(N^4 \, r_\mathrm{THC})$. Updates for the
rest of the terms in the THC decomposition can be calculated
similarly.

A simple iterative optimization algorithm can be built as follows.
First, the THC factors $W$ are initialized randomly.  For each factor,
an update is calculated as shown on
diagram~\ref{fig:least_squares_w1_sol}, keeping the other factors
fixed.  The process is iterated until convergence of the factors.  The
resulting THC-ALS algorithm is listed below.
\begin{algorithm}[H]
  \caption{Alternating Least Squares}\label{code:thc_als}
  \begin{algorithmic}[1] 
  \Function{thc-als}{$V, r_\mathrm{THC}, \epsilon$}
  \State $I_{1},I_{2},I_{3},I_{4} \gets$ size($V$)
  \State $W^1, W^2, W^3, W^4, X \gets$ init\_random($I_{1}, I_{2}, I_{3}, I_{4}, r_\mathrm{THC}$) 
  \Repeat \ForAll {$W \in \{W^1, W^2, W^3, W^4,
X\}$}
  \State $A_{W} \gets $ get\_environment($W^1, W^2, W^3, W^4, X$)
  \LineComment{$O(r_\mathrm{THC}^3)$}
  \State $B_{W} \gets $ get\_rhs($V,
   W^1, W^2, W^3, W^4, X$) \LineComment{$O(N^4 \, r_\mathrm{THC})$ or
$O(N^2 \, r_\mathrm{SVD} \, r_\mathrm{THC})$ with RI}
   \State $W_{new} \gets A^{-1} B$\Comment{$O(r_\mathrm{THC}^3)$}
   \EndFor
   \State $\Delta \gets \max_{W} \frac{\| W_{new} - W \|}{\|
W \|}$
   \State $W \gets W_{new}$ 
   \Until $\Delta > \epsilon$ 
   
   \Return $W^1, W^2, W^3, W^4, X$
    \EndFunction
  \end{algorithmic}
\end{algorithm}
The calculation of the right hand side of
Eq.~\ref{eq:least_squares_w1} dominates in the cost of THC-ALS,
scaling as $O(N^4 \, r_\mathrm{THC})$. A simple modification is
possible to reduce this cost by one order of magnitude.  If an
approximation to the singular vectors of the original tensor $V$ is
available from the beginning, as in the case of electron interaction,
it can be used in place of $V$, leading to a faster algorithm.  The
diagram corresponding to Eq.~\ref{eq:least_squares_w1} then becomes
\begin{equation}
\vcenter{\hbox{\includegraphics[height=20mm]{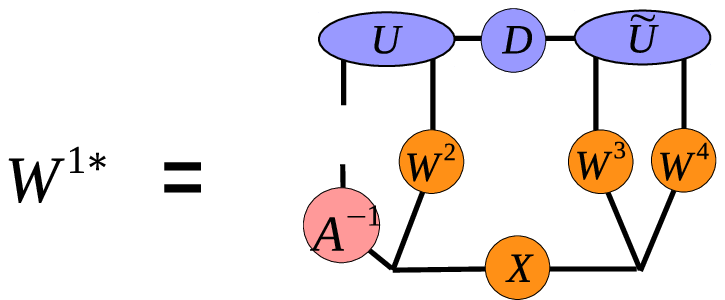}}}
\label{fig:least_squares_w1_sol_ri}
\end{equation}
The cost of the expression above scales as $O(N^2 \,
r_\mathrm{RI} \, r_\mathrm{THC})$, because the contraction of a
fourth-order tensor $V$ with matrices $W$ is replaced by contractions
of two third-order tensors $U$ and $\tilde{U}$.  We only need to
modify the function $get\_rhs()$ to build a lower scaling algorithm,
which we refer to as THC-ALS-RI.

Alternating least squares algorithms are simple and often
robust,\cite{uschmajew2012local} but may take a large number of
iterations to converge.\cite{comon2009tensor} Following an analogy
with CPD,\cite{sorber2013optimization}, we also implemented
quasi-Newton method using limited memory BFGS (L-BFGS) with a dogleg
trust region\cite{sorber2012unconstrained} for THC; this method we
refer to as THC-BFGS.

The THC-ALS and THC-BFGS, and their RI variants, are
novel direct methods to calculate the THC decomposition based on
minimization of the Frobenius norm of the error.  Composite methods
such as THC-CPD(ALS) and their RI variants have been used previously
in earlier work on THC.\cite{hohenstein_thc1}

We refer the reader to the supplementary material for
optimized expressions of the THC gradient and objective function. Due
to their complexity many of the equations we present (especially the
ones related to coupled cluster, see
Section~\ref{sec:tensor_structured_cc}) were generated by a computer
algebra system developed in our group,\cite{drudge1, drudge2} although
this can be done by manipulating diagrams as well.

\subsubsection{{Numerical Experiments}
\label{sec:numerical_experiments}} Here we wish to compare the
performance of the composite methods (THC-CPD(ALS), THC-CPD(BFGS),
THC-CPD(NLS)) and direct algorithms (THC-ALS, THC-BFGS) for THC
decomposition.  Table~\ref{tab:algorithms} shows the scaling per
iteration for the various algorithms we consider (see algorithms in
the text and also Ref.~\onlinecite{sorber2013optimization} for further
details on the scaling of CPD, which we used in the composite
methods).  The scaling is given for a full fourth-order tensor with
sizes equal $N$ in the first part of the table, and for RI-decomposed
tensors with rank $r_\mathrm{SVD}$ in the second part.  Recall that
the composite methods in the first part of the table require an
initial SVD, the cost of which scales as $O(N^4 \, r_\mathrm{SVD})$;
this cost is in addition to that of the iterative steps required to
converge the CPD.

\begin{table}[tb]
\caption{Computational scaling per iteration of various algorithms to
converge the CPD in composite methods or the THC itself in direct
methods.  The top half of the table shows scaling for methods which do
not use an initial RI, while the bottom half of the table shows
scaling for methods which do use an initial RI.
\label{tab:algorithms}}
\begin{ruledtabular}
 \begin{tabular}{lc} Algorithm & Scaling \\ \hline THC-CPD(ALS) &
$O(N^3 \, r_\mathrm{THC})$ \\ THC-CPD(BFGS) & $O(N^3 \,
r_\mathrm{THC})$ \\ THC-CPD(NLS) & $O(N^3 \, r_\mathrm{THC} +
r_\mathrm{THC}^3 + N^2 \, r_\mathrm{THC}^2)$ \\ THC-ALS & $O(N^4 \,
r_\mathrm{THC} + r_\mathrm{THC}^3)$ \\ THC-BFGS & $O(N^4 \,
r_\mathrm{THC} + r_\mathrm{THC}^3)$ \\ \hline THC-CPD-RI(ALS) & $O(N^3
\, r_\mathrm{THC})$ \\ THC-CPD-RI(BFGS) & $O(N^3 \, r_\mathrm{THC})$
\\ THC-CPD-RI(NLS) & $O(N^3 \, r_\mathrm{THC} + r_\mathrm{THC}^3 + N^2
\, r_\mathrm{THC}^2)$ \\ THC-ALS-RI & $O(N^2 \, r_\mathrm{SVD} \,
r_\mathrm{THC} + r_\mathrm{THC}^3)$ \\ THC-BFGS-RI & $O(N^2 \,
r_\mathrm{SVD} \, r_\mathrm{THC} + r_\mathrm{THC}^3)$
\end{tabular}
\end{ruledtabular}
\end{table}

To summarize the contents of Table~\ref{tab:algorithms}, let us assume
that both $r_\mathrm{SVD}$ and $r_\mathrm{THC}$ are $O(N)$, as is the
case for the electron interaction tensor.\cite{hohenstein_thc2} Then
all composite algorithms have a non-iterative $O(N^5)$ step followed
by iterative $O(N^4)$ steps, while direct algorithms have $O(N^5)$
cost per iteration.  If an RI approximation is used, all algorithms
have $O(N^4)$ scaling per iteration.

To get a feeling for how these various algorithms perform in practice,
we compared the convergence speed of direct and composite methods
using the performance metrics proposed by Dolan and
Mor\'e.\cite{dolan2002benchmarking} We generated fifty sets of random
THC factors using a uniform distribution, from which we built fifty
tensors which had size $4 \times 4 \times 4 \times 4$ and THC ranks 2
and 3.  We further generated fifty sets of random initial guesses
drawn from the same uniform distribution.  This yielded a set $P$ of
2500 (tensor, initial guess) pairs for each tensor rank.

\begin{figure*}[tb]
\includegraphics[width=0.95\columnwidth]{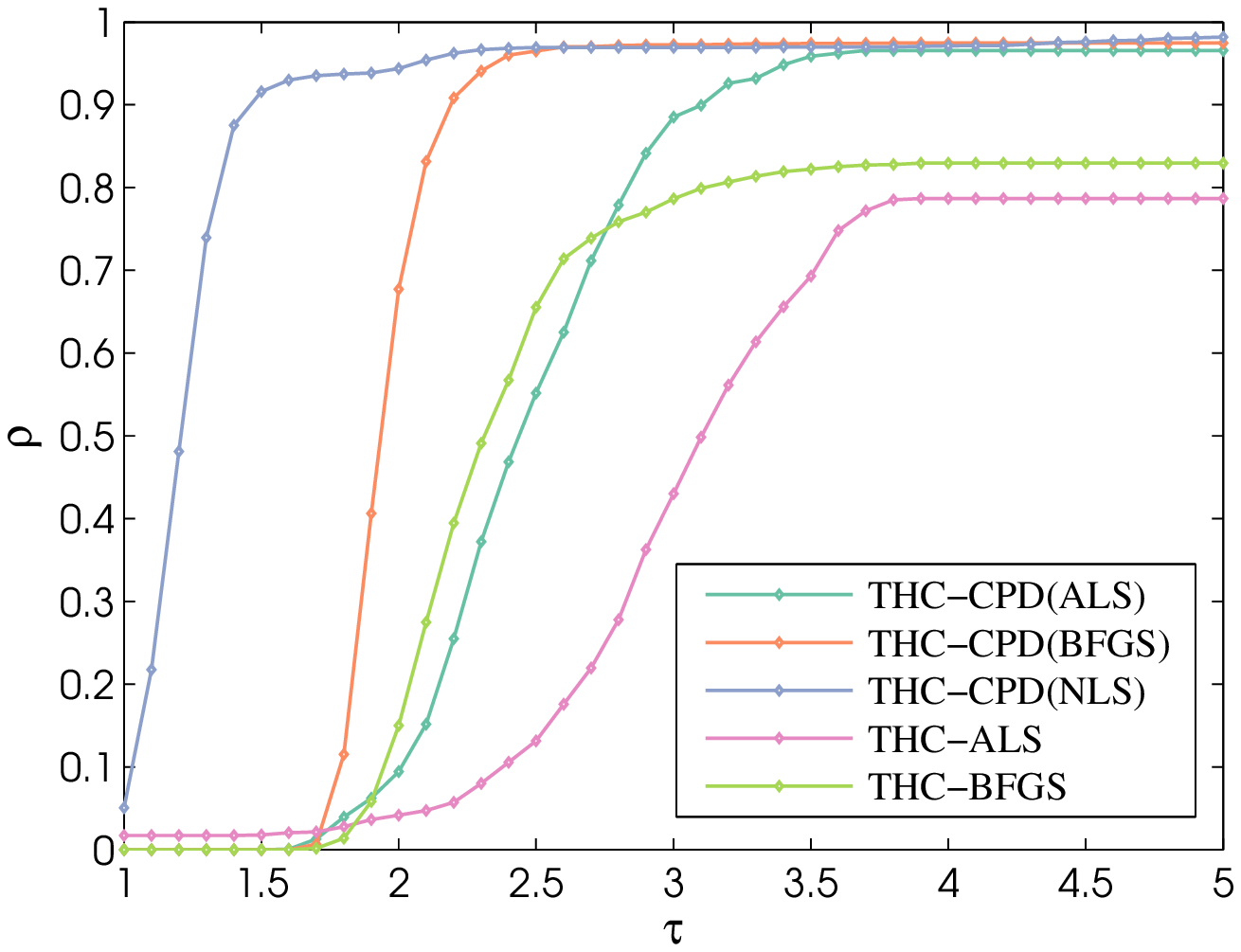}
\hfill
\includegraphics[width=0.95\columnwidth]{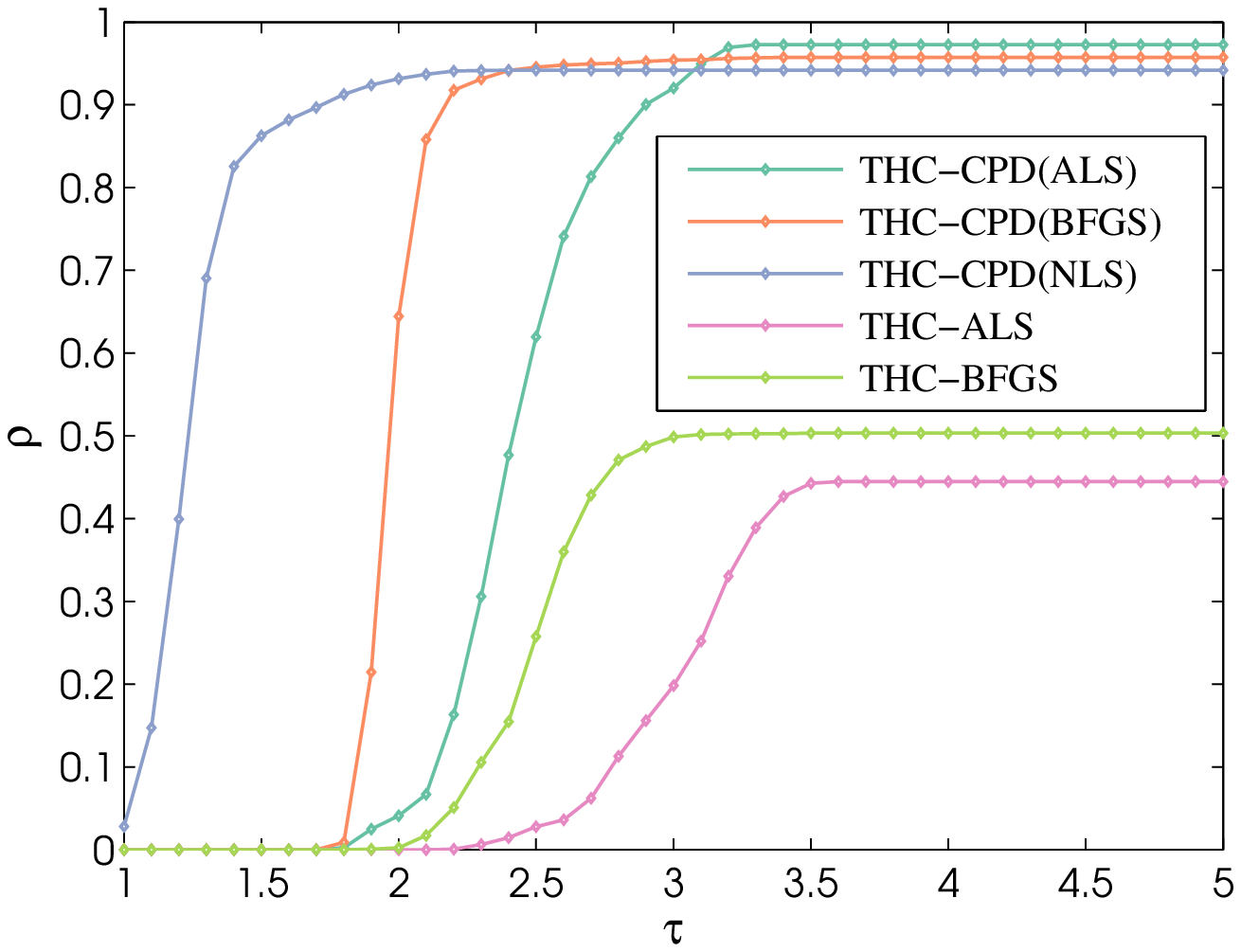}
\caption{Performance metric $\rho_s(\tau)$ for various THC
decomposition algorithms.  Left panel: $r_\mathrm{THC} = 2$.  Right
panel: $r_\mathrm{THC} = 3$.
\label{fig:alg_perf}}
\end{figure*}

The algorithms in the first part of Table~\ref{tab:algorithms}
(i.e. those algorithms that do not use RI) form a set of algorithms
$S$.  For each problem $p$ in $P$, we applied each algorithm $s$ in
$S$.  We allowed the algorithms to run for up to 2000 iterations or
until converged, where our convergence criterion was $\|V -
\tilde{V}\| \le 10^{-5}$.  The number of iterations required for an
algorithm $s$ to converge a problem $p$ we denote as $t_{p,s}$.  If an
algorithm did not converge a given problem, we set $t_{p,s}$ to
$\infty$.

For direct methods, we stopped the iterative algorithm if $\|V -
\tilde{V}\| \le 10^{-7} \, \|V\|$ and declared the method to have
failed if it did not meet our convergence threshold.  For composite
methods, we retained singular values larger than $10^{-7}$ in building
the factors $U_L$ and $U_R$.  We declared the CPD converged if $\|U -
\tilde{U}\| \le 10^{-10}$ and stopped the iterations if $\|U -
\tilde{U} \| \le 10^{-14} \, \| U \|$.  In all cases the threshold for
the pseudoinverse was set to $10^{-14}$.  We emphasize that for both
direct and composite methods the definition of success was accurate
decomposition of $V$, e.g. the magnitude of absolute error had to be
less than the threshold $\|V - \tilde{V}\| \le 10^{-5}$.

Having applied each algorithm $s$ to each problem $p$, we use as a
performance metric
\begin{equation} \rho_s(\tau) = \frac{| \{p \in P: t_{p, s} \leq
2^{\tau} \cdot \min_{s \in S}(t_{p, s}) \} | }{| P |}.
\end{equation} In other words, $\rho_s(\tau)$ is the fraction of
problems that algorithm $s$ solved within $2^\tau$ times the best
algorithm for each problem.  We would like $\rho_s(\tau)$ to approach
one for large enough $\tau$, indicating that the algorithm converged
all problems that could be converged, and we would like $\rho_s(\tau)$
to grow toward one as rapidly as possible, indicating that the
algorithm converged relatively quickly.  Results are shown in
Fig.~\ref{fig:alg_perf} where the left panel shows results for rank
two tensors and the right panel shows results for rank three tensors.

As one can see, composite methods THC-CPD outperform our direct THC
decomposition.  The difference in performance is more prominent for
$r_\mathrm{THC} = 3$ than it is for $r_\mathrm{THC} = 2$.  For
example, THC-ALS converges for less than $50\%$ of possible problems
when $r_\mathrm{THC} = 3$, compared to about $80\%$ for
$r_\mathrm{THC} = 2$.  We believe the poor performance of the direct
algorithm is because the THC factors are not unique (as our numerical
experimentation indicated), whereas the factors in the CPD are unique
under mild conditions.\cite{sidiropoulos2000uniqueness} This
non-uniqueness results in gradient vectors which are close to zero in
certain directions, and optimization algorithms then require many more
iterations to minimize the objective function.

Overall, the best method for THC seems to be the composite
THC-CPD(NLS), which we recall uses a nonlinear least squares solver
for CPD.\cite{sorber2013optimization,sorber2012unconstrained} We will
thus use THC-CPD(NLS) for subsequent THC decompositions in this work.

We should note that no method was able to solve all problems in our
setup, though the composite methods succeeded in the very large
majority of cases.  Similar behavior for random test factors was
previously observed for CPD.\cite{sorber2013optimization} This did not
pose a problem in our practical applications.  We should also note
that our results here should be considered with some caution, simply
because metrics generated with random factors may not be
representative for the tensors encountered in quantum chemistry, which
generally have more structure.  However, our results most likely show
the worst case behavior for the proposed algorithms.

\section{{Tensor Structured Coupled Cluster}
\label{sec:tensor_structured_cc}}
While the direct algorithms proposed in the previous
section were not particularly good for the decomposition
of random tensors, we introduced them because they find
new life in our tensor-decomposed coupled cluster methods,
as we discuss below. Let us begin, however, with a quick
overview of the restricted CCSD (RCCSD) method.  We define a cluster
operator
\begin{equation}
\hat{T} = {}^1\hat{T} + {}^2\hat{T},
\end{equation}
where the invidiual operators ${}^i\hat{T}$ are
excitation operators
\begin{subequations}
\begin{align} {}^1\hat{T} &= {}^1T_i^a \, \hat{E}_i^a, \\ {}^2\hat{T}
&= {}^2T_{ij}^{ab} \, \hat{E}_i^a \, \hat{E}_j^b.
\end{align}
\end{subequations}
Here,
\begin{equation}
\hat{E}_i^a = \hat{a}_{a,\uparrow}^\dagger \,
\hat{a}_{i,\uparrow} + \hat{a}_{a,\downarrow}^\dagger \,
\hat{a}_{a,\downarrow}
\end{equation}
are spin adapted excitations, or unitary group
generators,\cite{scuseria_ccsd} and ${}^iT$ are order $2i$ amplitude
tensors.  With these cluster operators, we construct a
similarity-transformed Hamiltonian $\bar{H}$ as
\begin{equation} 
\bar{H} = \mathrm{e}^{-\hat{T}} \, \hat{H} \,
\mathrm{e}^{\hat{T}},
\end{equation}
from which the energy can be extracted as
\begin{equation}
E_\mathrm{CCSD} = \langle 0| \bar{H} |0\rangle
\end{equation}
where $|0\rangle$ is a closed shell single determinant
(usually a Hartree-Fock state).  The excitation amplitudes are usually
obtained by projecting the similarity-transformed Hamiltonian on the
left against a set of excited determinants to form residuals which are
set to zero,
\begin{subequations}
\begin{align} {}^{1}R_{i}^{a} & = \langle 0|
\hat{a}_{i,\uparrow}^\dagger \, \hat{a}_{a,\uparrow} \, \bar{H} | 0
\rangle = 0, \\ {}^{2}R_{ij}^{ab} & = \langle 0|
\hat{a}_{i,\uparrow}^\dagger \, \hat{a}_{j,\downarrow}^\dagger \,
\hat{a}_{b,\downarrow} \, \hat{a}_{a,\uparrow} \, \bar{H} |0\rangle =
0.
\end{align}
\end{subequations}
These result in polynomial equations of the
amplitude tensors which can be transformed to the form
\begin{subequations}
\begin{align} {}^{1}T_{i}^{a} &= {}^{1}D_{i}^{a} ~ {}^{1}G_{i}^{a}, \\
{}^{2}T_{ij}^{ab} &= {}^{2}D_{ij}^{ab} ~ {}^{2}G_{ij}^{ab},
\label{eq:cc_amplitude_equations}
\end{align}
\end{subequations}
which can be solved by iterations until a fixed
point is found.  Here, ${}^1D$ and ${}^2D$ are orbital energy
denominator tensors built from diagonal elements of the Fock matrix
$F$:
\begin{subequations}
\begin{align} {}^1D_i^a &= \frac{1}{F_a^a - F_i^a}, \\
{}^{2}D_{ij}^{ab} &= \frac{1}{F_{a}^{a} + F_{b}^{b} - F_{i}^{i} -
F_{j}^{j}}.
\end{align}
\label{eq:denom_definition}
\end{subequations} 
The tensors ${}^1G$ and ${}^2G$ are built from
contractions of the amplitude tensors with the Hamiltonian.

\subsection{{Least Squares Coupled Cluster Theories}
\label{sec:least_squares_cc}} The logic used to derive the ALS
algorithm for THC decomposition can be readily applied in the coupled
cluster context.  Here, we will use coupled cluster doubles (for which
one neglects ${}^1\hat{T}$) as as example, with expressions for CCSD
shown in the supplementary material.

We begin by imposing the THC structure on the doubles amplitudes. We
approximate the amplitude tensor ${}^{2}T$ with its THC decomposition
${}^{2}\tilde{T}$.  The difference between original and approximated
amplitudes is
\begin{equation} 
\Delta_{T} = {}^{2}T - {}^{2}\tilde{T} = {}^{2}T -
(Y^{2} \odot Y^{1}) \cdot Z \cdot (Y^{4} \odot Y^{3})^{T},
\end{equation}
where $Y^{i}$ and $Z$ are factors in the THC
decomposition of ${}^{2}T$.  We wish to minimize the squared norm of
the error tensor $\Delta_{T}$, which is the minimization of the
corresponding cost function $f_{T}$,
\begin{equation}
f_{T} = |\Delta_{T}|^2 = ({}^{2}T^{\ast} -
{}^{2}\tilde{T}^{\ast})({}^{2}T - {}^{2}\tilde{T}).
\end{equation}
Setting partial derivatives of $f_{T}$ with respect to
the decomposition factors to zero, we obtain a new set of equations
\begin{equation} 
\frac{\partial f_{T}}{\partial Y} = - {}^{2}T^{\ast}
\frac{\partial {}^{2} \tilde{T}}{\partial Y} + {}^{2}\tilde{T}^{\ast}
\frac{\partial {}^{2} \tilde{T}}{\partial Y} = 0,
\end{equation}
where $Y \in \{Y^{1}, Y^{2}, Y^{3}, Y^{4}, Z\}$. Again,
as $f_{T}$ is real and analytic, only one set of derivatives (either
with respect to $Y$ or $Y^{\ast}$) is sufficient to find its minimum.

Now we use Eq. \ref{eq:cc_amplitude_equations} to replace ${}^2T^\ast$
with ${}^2D \, {}^2G$.  The idea is to thus to
minimize the difference between a decomposed tensor ${}^2\tilde{T}$
and a solution of the CCD amplitude equations.  The resulting
amplitude equations are
\begin{equation}
\tilde{T}^{\ast} \frac{\partial \tilde{T}}{\partial
Y} = {}^{2}G {}^{2}D \frac{\partial \tilde{T}}{\partial Y}.
\end{equation}
This is the analogue of Eq.~\ref{eq:least_squares_w1}
in THC-ALS, and can be solved in the least-squares sense (i.e. with
the help of the pseudoinverse) as the left-hand-side is linear in
$Y^\ast$.  Diagrammatically, we have
\begin{equation}
\vcenter{\hbox{\includegraphics[height=26mm]{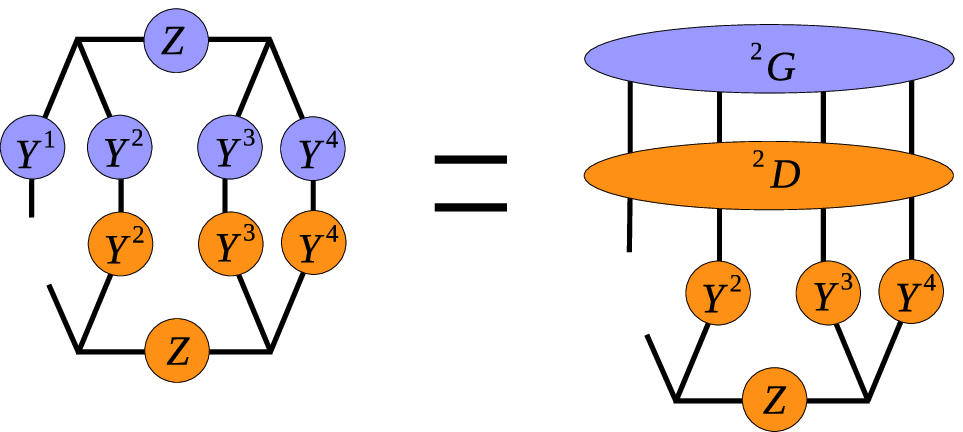}}}.
\label{fig:cc_thc}
\end{equation}
These equations can be further factorized if one employs CPD of
${}^2D$ to disentangle particle and hole indices.  A low-rank
decomposition of denominator tensors can be built using an exponential
parametrization\cite{braess2005approximation} (also known as Laplace
transformation)\cite{almlof1991elimination} as, for example,
\begin{subequations}
\begin{align} {}^2D_{ij}^{ab} &= C_\omega \, \mathrm{e}^{A_\omega \,
F_i^i} \, \mathrm{e}^{A_\omega \, F_j^j} \, \mathrm{e}^{-A_\omega \,
F_a^a} \, \mathrm{e}^{-A_\omega \, F_b^b} \\ &= {}^2D^1_{i,\omega} \,
{}^2D^2_{j,\omega} \, {}^2D^3_{a,\omega} \, {}^2D^4_{b,\omega}.
\end{align}
\end{subequations}
We have used the parameters from
Ref.~\onlinecite{braess2005approximation}, which provide absolute
accuracy of better than $10^{-12}$ with ranks of order $\approx 15$,
which do not depend on the system size $N$.

The final form of our ALS-type coupled cluster doubles equations is
thus
\begin{equation}
\vcenter{\hbox{\includegraphics[height=30mm]{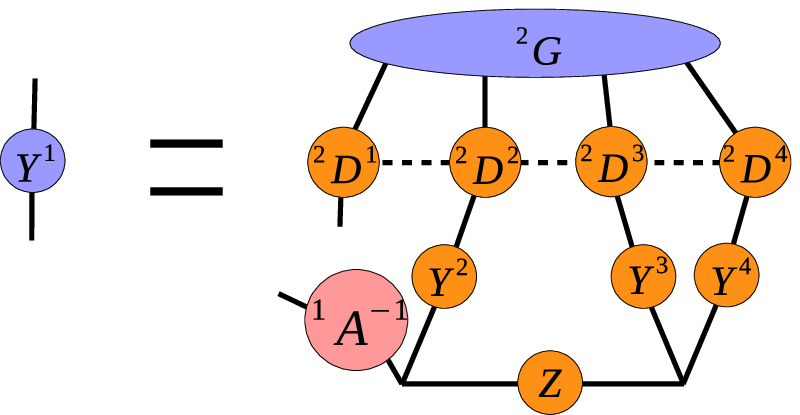}}}.
\label{fig:cc_thc_als}
\end{equation}
The explicit form of these equations
and analogous expressions for ALS-type CCSD are shown in the
supplementary material.  After defining proper intermediates, which we
did using our automatic algebraic system,\cite{drudge2} the cost of
these equations has quartic scaling in $r_\mathrm{THC}$ and $N$ per
iteration.  We provide those fully factorized equations in the
supplementary material along with the source code for the
contractions. Most of the numerical experiments in the following
section were done with a simpler code which had $O(N^5)$ scaling
because it made less sophisticated use of
intermediate quantities; however, the $O(N^4)$ and $O(N^5)$
implementations differ only in the order in which contractions were
carried out.

Equation \ref{fig:cc_thc_als} and its analogs for all
other factors in the decomposition of ${}^2T$ constitute what we call
THC-RCCSD and are the main result of this paper.  It must be stressed
that the proposed scheme is generic, and can be applied to any
factorization of amplitudes and the Hamiltonian.  We use THC here, and
leave the exploration of other possibilities for subsequent work.

\subsection{{Test Calculations} \label{sec:test_calculations}} To
assess the performance of our tensor-structured CCSD, we present
calculations on a variety of small- to medium-sized molecules.  All
calculations used the cc-pVDZ basis from EMSL
database,\cite{schuchardt2007basis} and the corresponding cc-pVDZ-RI
was used in the RI approximation.

For smaller systems the THC-CPD(NLS) algorithm was used to obtain the
THC approximation to the full two-electron integrals in the AO basis.
We set the relative convergence threshold for CPD iterations to
$10^{-14}$, as we did in our numerical experiments in
Sec.~\ref{sec:tensor_decompositions}.  Singular values larger than
$10^{-12}$ were retained in obtaining $U_L$ and $U_R$.  The maximum
number of iterations allowed during the decomposition of the integrals
was 1000.  The subsequent coupled cluster calculations were stopped
either after the energy was converged to within $10^{-9}$ Hartree or a
limit of 200 iterations was reached.  Thresholds for pseudoinverse
calculations were set to $10^{-14}$.

For larger systems, listed in Tab.~\ref{Tab:Energies}, THC-CPD(NLS)
was applied to RI-decomposed two-electron integrals.  Other parameters
were as described above, except we decreased the number of iterations
allowed during decomposition of the integrals to 500 and the number of
coupled cluster iterations allowed to 100.

\subsubsection{{Decomposition of Two-Electron Integrals}
\label{sec:decomposition_of_two_electron_integrals}} 
\begin{figure}[tb]
\includegraphics[width=\columnwidth]{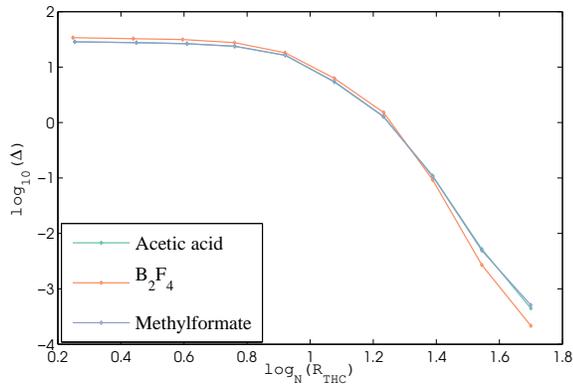}
\caption{Frobenius norm of error in decomposed two electron integrals.
\label{fig:thc_err_mo_3systems}}
\end{figure}

The accuracy of
the THC decomposition of the two-electron integrals governs the
accuracy of the energy in subsequent calculations.  Thus, we first
wish to check the dependence on the error in the decomposition of
two-electron integrals on THC rank.
Figure~\ref{fig:thc_err_mo_3systems} plots this error in a double
logarithmic scale for three small molecules.  We note that the
decomposition is computationally useful if the rank $r_\mathrm{RHC}$
is close to the number of basis functions $N$.  As the figure shows,
the error in the two-electron integrals decays exponentially with
respect to THC rank.  We found that this trend holds for every system
tested and depends only slightly on whether the two-electron integrals
are decomposed in the atomic orbital or molecular orbital basis.

\begin{figure}[tb]
\includegraphics[width=\columnwidth]{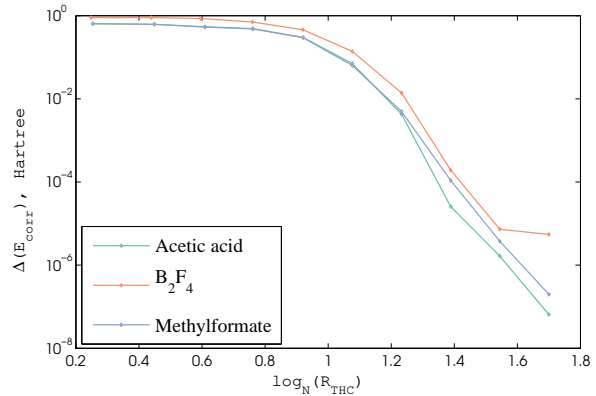}
\caption{Absolute error in the MP2 correlation energy.
\label{fig:mp2_err_ao_full}}
\end{figure}

To see how the decomposition affects subsequent energies, we checked
the error in the second-order M{\o}ller-Plesset (MP2) correlation
energy, as shown in Fig.~\ref{fig:mp2_err_ao_full}.  The combination
of MP2 and THC was first proposed by Hohenstein \emph{et
al}.\cite{hohenstein_thc2} and scales as $O(N^4)$.  These authors used
a version of THC with the restriction that all factors $W$ were the
same, which we did not impose in our work. 
The error in the MP2 correlation energy follows the trend seen in the
decomposition of the two-electron integrals.  Results within $0.1~mH$
of the exact MP2 correlation energy are already achieved with
$r_\mathrm{THC} \sim N^{1.2} - N^{1.4}$.  We expect that the THC would
work better for larger and more extended systems as the two-electron
integrals become sparser and a lower rank decomposition would
correspondingly become more accurate.

\subsubsection{{Restricted Coupled Cluster with Singles and Doubles}
\label{sec:restricted_coupled_cluster_with_singles_and_doubles}}
\begin{figure}[tb]
\includegraphics[width=\columnwidth]{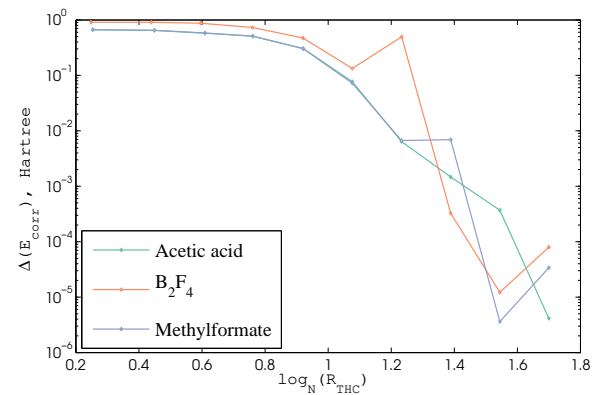}
\caption{Absolute error in the RCCSD correlation energy.
\label{fig:cc_err_ao_full}}
\end{figure}

\begin{figure}{tb}
\includegraphics[width=\columnwidth]{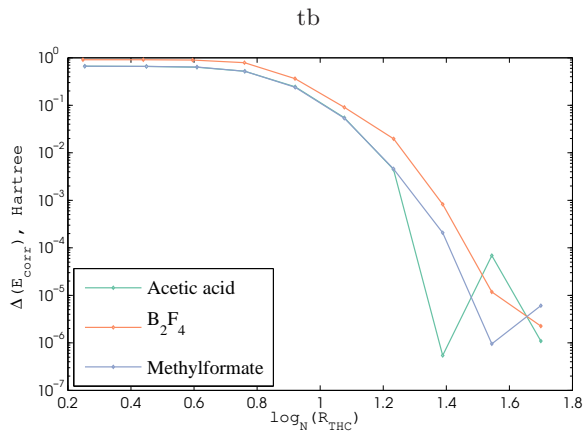}
\caption{Absolute error in the RCCSD correlation energy with exact two
electron integrals.
\label{fig:cc_err_ao_full_amps_only}}
\end{figure}

Finally, we demonstrate the behavior of the THC-decomposed RCCSD
method (THC-RCCSD), seen in Fig.~\ref{fig:cc_err_ao_full}.  We chose
the rank of the THC decomposition of the amplitudes and two-electron
integrals to be the same.  The error in the RCCSD correlation energy
has a non-monotonic dependence on THC rank, but follows the same basic
trends as seen in Fig.~\ref{fig:thc_err_mo_3systems} and
Fig.~\ref{fig:mp2_err_ao_full}.  As with MP2, errors on the order of
$0.1~mH$ are achieved with $r_\mathrm{THC} \sim N^{1.2} - N^{1.4}$.
It is interesting to see what part of the error in energy can be
attributed to the approximation of the Hamiltonian, especially because
building the decomposition of the Hamiltonian contributed $\sim 95\%$
of the total computational cost.  For this reason we calculated the
correlation energy with converged THC-RCCSD amplitudes but exact
two-electron integrals.  As Fig.~\ref{fig:cc_err_ao_full_amps_only}
shows, using the exact two-electron integrals decreases the error in
energy, as one would expect, but does not remove its non-monotonic
dependence on the THC rank.  We attribute this behavior to the
nonlinear nature of the coupled cluster equations, which can be quite
sensitive to changes in the parameters of the Hamiltonian.

Having seen how the THC-RCCSD method performs for various THC
decomposition ranks, we tested the method on a set of small and
medium-sized molecules introduced in previous work on
THC.\cite{hohenstein_thc3} Technical details of the calculations,
including molecular geometries and reference energies, are provided in
the supplementary materials.  We chose the ranks of the THC
decomposition of the amplitudes and integrals to be similar to the
number of functions $N_\mathrm{RI}$ in the basis used in the RI
approximation.  Results are presented in Table~\ref{Tab:Energies}.  We
used RI for all these calculations.
\begin{table}[t]
\caption{CCSD correlation energies ($E_c$) and errors in the THC-RCCSD
correlation energies ($\Delta E_c$) for several small molecules.
\label{Tab:Energies}}
\begin{ruledtabular}
\begin{tabular}{lccc}
& & \multicolumn{2}{c}{$\Delta E_c (mH)$} \\
\cline{3-4} System & $E_c (mH)$ & $N_\mathrm{THC} = N_\mathrm{RI}$ &
$N_\mathrm{THC} = 1.5 \, N_\mathrm{RI}$ \\
\hline
Acetic acid & -666.510 & -0.579 & -0.453 \\
Aniline & -997.193 & -1.177 & -0.471 \\
Diboron tetrafluoride & -909.944 & -0.702 & -0.716 \\
Benzene & -823.101 & -0.985 & -0.450 \\
Butadiene & -581.340 & -0.710 & -0.274 \\
Cyclobutane & -621.099 & -0.895 & -0.290 \\
Dimethylsulfoxide & -661.870 & 0.195 & -0.624 \\
Furan & -736.463 & -0.865 & -0.454 \\
Isobutane & -652.505 & -0.876 & -0.263 \\
Methylformate & -666.805 & -0.586 & -0.455 \\
Methylnitrite & -708.990 & -0.476 & -0.492 \\
Phenol & -1005.727 & -0.887 & -0.514 \\
Pyridine & -842.453 & -1.045 & -0.475 \\
Pyrrole & -727.051 & -0.855 & -0.407 \\
Thiophene & -695.593 & -1.013 & -0.657 \\
Toluene & -980.030 & -1.270 & -0.461 \\
\hline
MUE\footnote{mean unsigned error} & & 0.820 & 0.466\\
Max\footnote{maximum unsigned error} & & 1.270 & 0.716\\
RMS\footnote{root-mean-square error} & & 0.861 & 0.482\\
\end{tabular}
\end{ruledtabular}
\end{table}
We note that our results are on par with calculations of
Hohenstein \emph{et al.},\cite{hohenstein_thc3} but similar errors are achieved with
ranks which are roughly half as large.  Presumably this is because in
previous work most of the factors in the THC decomposition of the
amplitudes were kept fixed (except $Z$), whereas our scheme optimizes
all factors, therefore providing greater flexibility and reaching the
exact decomposition faster.  Again, we emphasize that the proposed
scheme is not limited to THC, and can be applied to many other
decompositions, which is the topic of ongoing investigation.

\section{Conclusions}
Systematically dependable quantum chemical
methods rely on solving the Schr\"odinger equation, but unfortunately
do so at a significant and often impractical computational cost.  For
many-body methods such as coupled cluster theory, the cost can be
explained simply: the various objects of the theory are high-order
tensors which must be contracted with one another, and the contraction
of two high-order tensors is computationally costly.  Tensor
decompositions lower the cost by writing high-order tensors as sums of
products of low-order objects, and are one of the most promising ways
to apply many-body theories to large systems.

In this work, we have shown how the combination of tensor
hypercontraction and canonical polyadic decomposition allows us to
solve the closed-shell CCSD equations with $O(N^4)$ scaling by solving
directly for the factors which decompose the cluster operator
(Eqn.~\ref{fig:cc_thc_als}).  By increasing the
dimensions of these factors (i.e. by increasing the rank) we can
approach the exact CCSD result in a more or less systematic fashion,
and can achieve results within $0.1 mH$ of the exact CCSD answer with
ranks on the order of the size of the basis.  Our alternating least
squares method improves over previous studies of THC in coupled
cluster theories\cite{hohenstein_thc3, hohenstein_cc2} where fixed
real-space quadratures were used to build the decomposition of cluster
amplitudes and provides more accurate results for smaller
ranks. The proposed scheme, however, is general and
can be applied to any decomposition, as well as readily extended to
more sophisticated coupled cluster theories. Among other
possibilities, we plan extensions to the Unrestricted CC and our own
symmetry-projected CC theories.\cite{qiu2017projected1,
qiu2017projected2, gomez2017attenuated} Lastly, we should mention
that coupled cluster methods with decomposed amplitudes are much more
suitable for parallelization than are the traditional ones, because
the communication becomes much cheaper. While our work along the
mentioned lines is still in the early stages, we hope that these
low-scaling coupled cluster methods will help make large-scale CCSD
calculations essentially routine.

\section*{Supplementary material} See supplementary materials for the
THC gradient expressions, complete specification of test systems and
least squares coupled cluster expressions.

\begin{acknowledgments} This work was supported as part of the Center
for the Computational Design of Functional Layered Materials, an
Energy Frontier Research Center funded by the U.S. Department of
Energy, Office of Science, Basic Energy Sciences under Award \#
DE-SC0012575. G.E.S. is a Welch Foundation chair (C-0036).
\end{acknowledgments}

\appendix
\section{{Wiring Diagrams}
\label{sec:appendix_diagrams}}
We have made extensive use of wiring
diagrams to simplify the representation and manipulation of complex
tensor expressions.  This graphical notation is similar to the usual
diagrammatic notation used in many-body theory, but not identical.
For completeness, we here describe the basic semantics of our
diagrams.

In our notations, tensors are represented by shapes.  Typically a
$d$-order tensor is represented by a polygon with $d$ corners (and a
second-order tensor by a circle), though we have not followed this
convention universally.  Indices are denoted by lines; a line
connecting multiple tensors is to be summed over, and open lines
correspond to free indices.  If a particular element of a tensor
expression is required, we label the open lines.

To be concrete, a matrix product would be represented by
\begin{equation}
\vcenter{\hbox{\includegraphics[height=8mm]{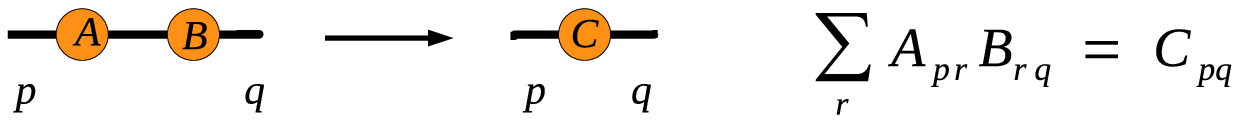}}}
\end{equation}
and a more general contraction of a fourth-order tensor
with a third-order tensor can be drawn as
\begin{equation}
\vcenter{\hbox{\includegraphics[height=8mm]{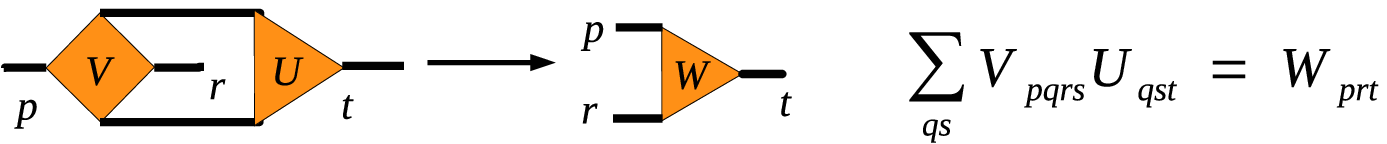}}}.
\end{equation}
Diagrams can be used to readily estimate the cost of contractions (and
other operations).  The cost $\Omega$ of contracting two tensors over
$L$ indices of size $\{ \lambda \}_1^{L}$ to a tensor with $M$ indices
of size $\{ \mu \}_1^M$ scales with respect to $N$ as
\begin{equation} 
\Omega = \mathcal{O}(N^{\sum_{l = 1}^{L} log_{N}
\dim(\{\lambda\}_l) \cdot \sum_{m = 1}^M log_{N} \dim(\{\mu\}_m)}).
\label{eq:contract_scaling_estimate}
\end{equation} 
One can simply estimate the scaling of a contraction by
multiplying the dimensions of each open line in the result together
with those of each closed line.  For example, a contraction of two
third-order tensors of size $N \times N \times N$ over two indices of
size $N$ scales as $O(N^4)$:
\begin{equation}
\vcenter{\hbox{\includegraphics[height=12mm]{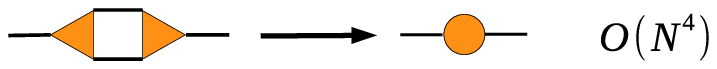}}}
\label{fig:contraction_scaling}
\end{equation}
Other operations that can be represented pictorially are of an outer
product type. This situation corresponds to merging the nodes together
and leaving all lines in the final structure:
\begin{equation}
\vcenter{\hbox{\includegraphics[height=18mm]{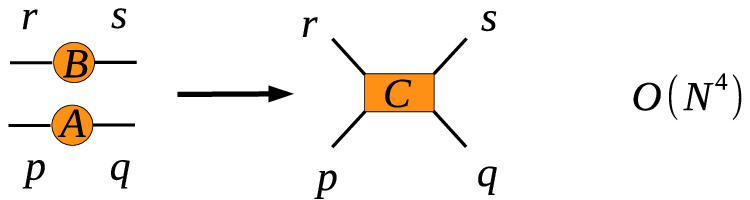}}}
\label{fig:outer_product}
\end{equation} 
Note that if one reshapes the fourth-order tensor above
into a matrix with combined indices $rp$ and $sq$, then the result
will coincide with the usual Kronecker product of matrices, where we
recall that the Kronecker product is
\begin{equation}
C = A \otimes B \Leftrightarrow C_{rp, sq} = A_{p,q}
\cdot B_{r,s}.
\end{equation}
The cost of product-type operations is
\begin{equation}
\Omega = \mathcal{O}(N^{\sum_{m = 1}^{M} log_{N}
dim(\{\mu\}_m)})
\label{eq:outer_scaling_estimate}
\end{equation}
where $\{ \mu \}_1^{M}$ are $M$ free indices in the
resulting tensor.

For our purposes we slightly extended the diagrammatic notation by
introducing summations over an index shared by more than two terms. We
denote such indices by branching lines with a dot at the branching
point. This dot can be interpreted either as an index of the summation
itself, or as a fully diagonal tensor whose elements are contractions
of Kronecker deltas, e.g.
\begin{equation} K_{p,q,r,\ldots} = \sum_\alpha \delta_p^\alpha
\delta_q^\alpha \delta_r^\alpha \ldots.
\end{equation}
The latter interpretation means that all contractions
in the diagrams can be thought pairwise as in the normal case.
Although not quite standard, this extension has been used before in
the tensor network literature.\cite{ying2016tensor} Using our new
notation, contracting a canonical polyadic decomposition of a third
order-tensor back to a full tensor can be denoted as
\begin{equation}
\vcenter{\hbox{\includegraphics[height=28mm]{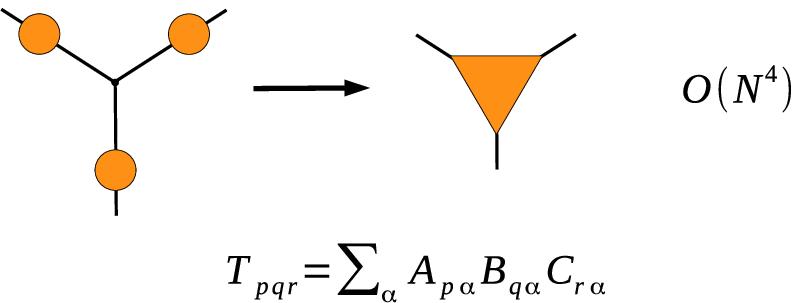}}}
\label{fig:dot_in_diagrams}
\end{equation}
If the dimensions of this tensor are $N \times N \times
N$ and the rank of the decomposition (the dimension of the auxiliary
index $\alpha$) is $N$, then the cost of rebuilding the original
tensor from its decomposed form will scale as $O(N^4)$. We note that
Eq.~\eqref{eq:contract_scaling_estimate} holds in this case just the
same way as with normal pairwise contractions.

Let us also list diagrammatic representations of common matrix
operations.  The Frobenius norm of a tensor, which we recall is
\begin{equation} \| A \| = \sqrt{\sum_{p} \sum_{q} \sum_{r} \ldots
A_{pqrs\ldots} \, A^\ast_{pqrs\ldots}},
\end{equation}
is given diagrammatically as the square root of a
tensor fully contracted with its own conjugate:
\begin{equation}
\vcenter{\hbox{\includegraphics[height=10mm]{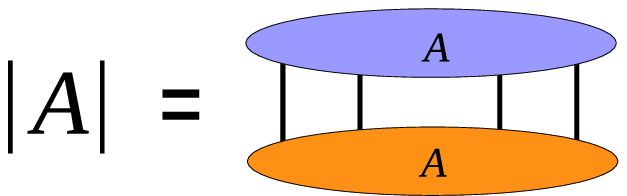}}}
\label{fig:frobenius_norm}
\end{equation}
We have used a darker color to denote complex
conjugation here.

The column-wise Khatri-Rao product is
\begin{equation}
D = A \odot B ~~\Leftrightarrow ~~D_{qp,\alpha} =
A_{p,\alpha} \cdot B_{q,\alpha}.
\end{equation}
Note that $A$ and $B$ should have the same number of
columns to be compatible.  The resulting matrix $D$ can be reshaped to
a third-order tensor with indices $p$, $q$ and $\alpha$.
Diagrammatically, the Khatri-Rao product is
\begin{equation}
\vcenter{\hbox{\includegraphics[height=18mm]{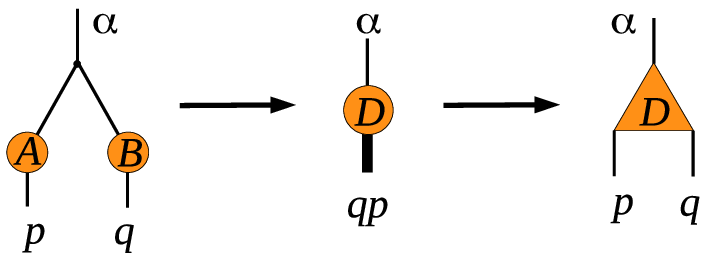}}}
\label{fig:khatri_rao_product}
\end{equation} 
Here we used a thick line to denote a combined index
$qp$. Note also that the canonical polyadic decomposition can be
conveniently expressed through the Khatri-Rao product, which is also
reflected by the diagrams:
\begin{equation}
\vcenter{\hbox{\includegraphics[height=40mm]{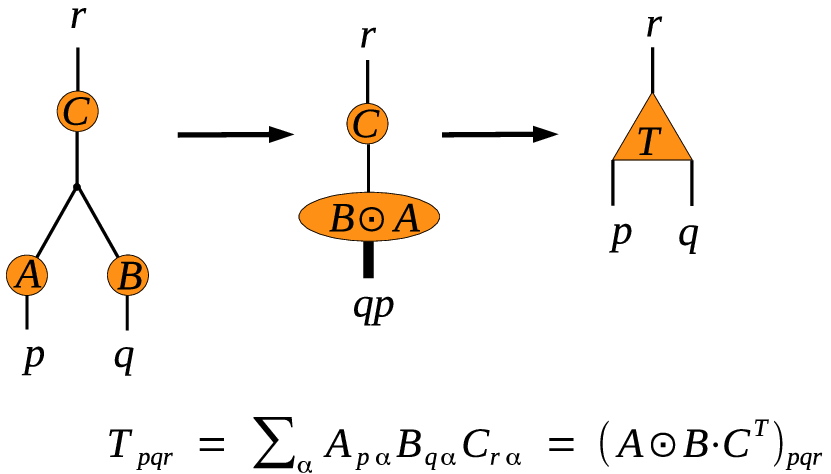}}}.
\label{fig:canonical_khatri_rao}
\end{equation}
Finally, we point out that wiring diagrams provide an easy way to
calculate derivatives.  A partial derivative of a tensor network with
respect to one of its component tensors is simply the network with
that tensor removed.

\bibliography{references}
\end{document}